\documentclass[conference]{IEEEtran}
\usepackage[table,xcdraw,usenames,dvipsnames]{xcolor}
\usepackage{array}
\usepackage{float}
\usepackage{graphicx}
\usepackage[english]{babel}
\usepackage{pgfplots}
\usepackage{amsmath}
\usepackage{booktabs}
\usepackage{multirow}
\usepackage{array,booktabs,arydshln,xcolor}

\usepackage{tikz}
\usepackage[table,xcdraw]{}
\usepackage{float}

\begin{document}

\title{Dynamic Environments for Virtual Machine Placement considering Elasticity and Overbooking}
\author{\IEEEauthorblockN{Jammily Ortigoza}
\IEEEauthorblockA{Science and Technology School\\Catholic University of Asunci\'on\\jortigozaf@gmail.com\\Paraguay}
\and
\IEEEauthorblockN{Fabio L\'opez-Pires}
\IEEEauthorblockA{Itaipu Technological Park\\National University of Asunci\'on\\fabio.lopez@pti.org.py\\Paraguay}
\and
\IEEEauthorblockN{Benjam\'in Bar\'an}
\IEEEauthorblockA{National University of Asunci\'on\\Catholic University of Asunci\'on\\bbaran@pol.una.py\\Paraguay}
}

\maketitle

\begin{abstract}
Cloud computing datacenters provide millions of virtual machines in actual cloud markets. In this context, Virtual Machine Placement (VMP) is one of the most challenging problems in cloud infrastructure management, considering the large number of possible optimization criteria and different formulations that could be studied. Considering the on-demand model of cloud computing, the VMP problem should be solved dynamically to efficiently attend typical workload of modern applications. This work proposes a taxonomy in order to understand possible challenges for Cloud Service Providers (CSPs) in dynamic environments, based on the most relevant dynamic parameters studied so far in the VMP literature. Based on the proposed taxonomy, several unexplored environments have been identified. To further study those research opportunities, sample workload traces for each particular environment are required; therefore, basic examples illustrate a preliminary work on dynamic workload trace generation.
\end{abstract}
\IEEEpeerreviewmaketitle

\hyphenation{re-pre-sent taxo-no-my pro-blem con-so-li-da-ted appro-xi-ma-ting sce-na-rios re-pre-sents accor-ding pro-ce-ssing fo-llo-wing con-si-de-ring vir-tua-li-zed ne-ce-ssa-ry en-for-cing re-so-lu-tion pre-fe-ren-ces co-rres-ponds re-pre-sen-ting E-las-ti-ci-ty mo-ni-tor cha-rac-te-ri-za-tion re-co-mmen-ded addre-sing exam-ple classi-fi-ca-tion pro-fi-ta-ble over-boo-king par-ti-cu-lar fe-de-ra-ted Addi-tio-na-lly com-ple-xi-ty me-mo-ry a-ppli-ca-tions cha-llen-ges ge-ne-ra-tor asso-cia-ted pu-bli-shed}

\section{Introduction}
\label{introduction}
{\setlength{\parindent}{12pt}	

The rapid demand growth for computational resources in modern business and scientific applications presents several challenges for design, implementation and management of scalable datacenters to meet the requirements of customers in a competitive and efficient way \cite{soundararajan2010challenges}. 

Considering the evolution of resource provisioning, three main models could be identified: (1) traditional provisioning of resources with independent physical hardware, (2) modern provisioning of shared resources through virtualized hardware and (3) trending dynamic provisioning of resources through a cloud computing model \cite{buyya2008market}. The traditional provisioning environment has mostly evolved to a virtualized provisioning of resources in current datacenters, considering its advantages for management and resource utilization.

Virtualization in modern datacenters introduces complex management decisions related to the placement of virtual machines (VMs) into the available physical machines (PMs). In this context, Virtual Machine Placement (VMP) represents the process of selecting which VMs should be executed in a given set of PMs of a datacenter \cite{lopez2015b}. The VMP problem is mostly formulated as a combinatorial optimization problem, representing one of the most challenging problems in virtualized datacenters infrastructure management, considering the large number of possible optimization criteria and different formulations that could be studied \cite{lopez2015}.

For virtualized datacenters with deployments of VMs that rarely change its configuration over time, a static (offline) formulation of the VMP problem may be appropriate \cite{lopez2013virtual}. Additionally, in virtualized datacenters where a small number of VMs are created and destroyed, a semi-static formulation of the VMP problem could be acceptable (e.g. consolidating VMs every day at midnight). On the other hand, considering the today more realistic on-demand model of cloud computing with dynamic resource provisioning, a static (or semi-static) formulation of the VMP problem can result in under-optimal solutions after a short period of time. Clearly, the VMP problem for cloud computing environments must be formulated as a pure dynamic (online) optimization problem to efficiently attend dynamic workload of modern applications \cite{lopez2015}.

\subsection{Background and Motivation}
\label{motivation}
The VMP problem has been extensively studied and several surveys have already been presented in the VMP literature. Existing surveys focus on specific issues such as: (1) energy-efficient techniques applied to the problem \cite{beloglazov2012energy,salimian2013survey}, (2) particular architectures where the VMP problem is applied, as federated clouds \cite{gahlawat2014survey}, and (3) methods for comparing performance of placement algorithms in large on-demand clouds \cite{mills2011comparing}. None of the mentioned surveys presented a general and extensive study of a large part of the VMP literature. In consequence, L{\'o}pez-Pires and Bar{\'a}n presented in \cite{lopez2015} an extensive up-to-date survey of the most relevant VMP literature and proposed a novel taxonomy in order to identify research opportunities defining a general vision on this problem.

According to \cite{lopez2015}, the VMP problem is mostly formulated as an online optimization problem, where live migration techniques allow VMs to be dynamically consolidated on necessary PMs according to dynamic requirements of resources. The most studied environment for online formulations of the VMP problem considers that VMs are dynamically created and destroyed \cite{lopez2015}. To the best of the authors' knowledge, there is no published work presenting a detailed characterization of possible dynamic environments for the VMP problem.

Clearly, a deeper research of possible dynamic parameters in cloud computing is necessary in order to propose holistic and more realistic environments for the formulation of the VMP problem for cloud computing datacenters. 

Consequently, this work complements the taxonomy presented by the authors in \cite{lopez2015} focusing specifically on dynamic formulations of the VMP problem from the providers' perspective, proposing a taxonomy in order to understand possible challenges for Cloud Service Providers (CSPs) in dynamic environments to efficiently attend customers' requests for virtual resources, based on the most relevant dynamic parameters studied so far in the VMP literature. The taxonomy proposed in this work must be jointly studied with the taxonomy first proposed in \cite{lopez2015} in order to represent a complete VMP problem.

The remainder of this paper is organized as follows: Section \ref{literature} details the literature selection process considered in this work, while Section \ref{criteria} introduces the classification criteria of the proposed taxonomy. Section \ref{taxonomy} presents the proposed taxonomy detailing the mathematical notation and basic examples of the identified dynamic environments for the VMP problem. Based on the proposed taxonomy, Section \ref{workloads} presents a preliminary work on generation of workload traces for the identified dynamic environments. Finally, conclusions and future work are left to Section \ref{conclusions}.

\section{Reviewed Literature}
\label{literature}

\subsection{Keywords Search}
The selection process of relevant articles started with a search for research articles from Google Scholar database [scholar.google.com] with at least one of the following selected keywords in the article title: (1) virtual machine placement, (2) vm placement, (3) virtual machine consolidation, (4) vm consolidation or (5) server consolidation.

\subsection{Publisher Filtering}
Considering the large number of results from keywords search step, the literature selection process focused on research articles from the following well-known publishers: (1) ACM, (2) IEEE, (3) Elsevier and (4) Springer. This filtering step resulted in a reduction from 446 to 172 research articles. A detailed list of the 172 resulting articles can be found in \cite{lopez2014survey}.

\subsection{Abstract Reading}
Considering the 172 resulting articles from the publisher filtering step, a reading of the abstracts was performed in order to identify the most relevant articles that specifically study the VMP problem. Additionally, short papers (i.e. research articles with less than 6 pages) were removed from the selected literature, resulting in 84 selected articles of the VMP literature. A detailed list of the 84 resulting articles can be found in \cite{lopez2014survey}.

\subsection{Online Formulations for Provider-oriented VMP Problem}
Based on the  84 studied articles addressed in \cite{lopez2015}, this work selected the 64 articles that proposed online formulations for the VMP problem from the providers' perspective, considering the relevance of this type of environments for actual cloud computing providers. An in-depth reading of this universe of 64 articles was performed with the aim of identifying the most relevant dynamic parameters.

\section{Classification Criteria}
\label{criteria}
This work identified the following dynamic parameters: 

\begin{itemize}	
\item resource capacities of VMs (vertical elasticity);
\item number of VMs of a service (horizontal elasticity);
\item utilization of resources of VMs (related to overbooking).
\end{itemize}

Consequently, dynamic environments for online formulations of the provider-oriented VMP problem may be classified by one or more of the following classification criteria: \hspace{1cm} (1) elasticity and (2) overbooking.

\begin{figure}[b]
\centering
\includegraphics[scale=0.45]{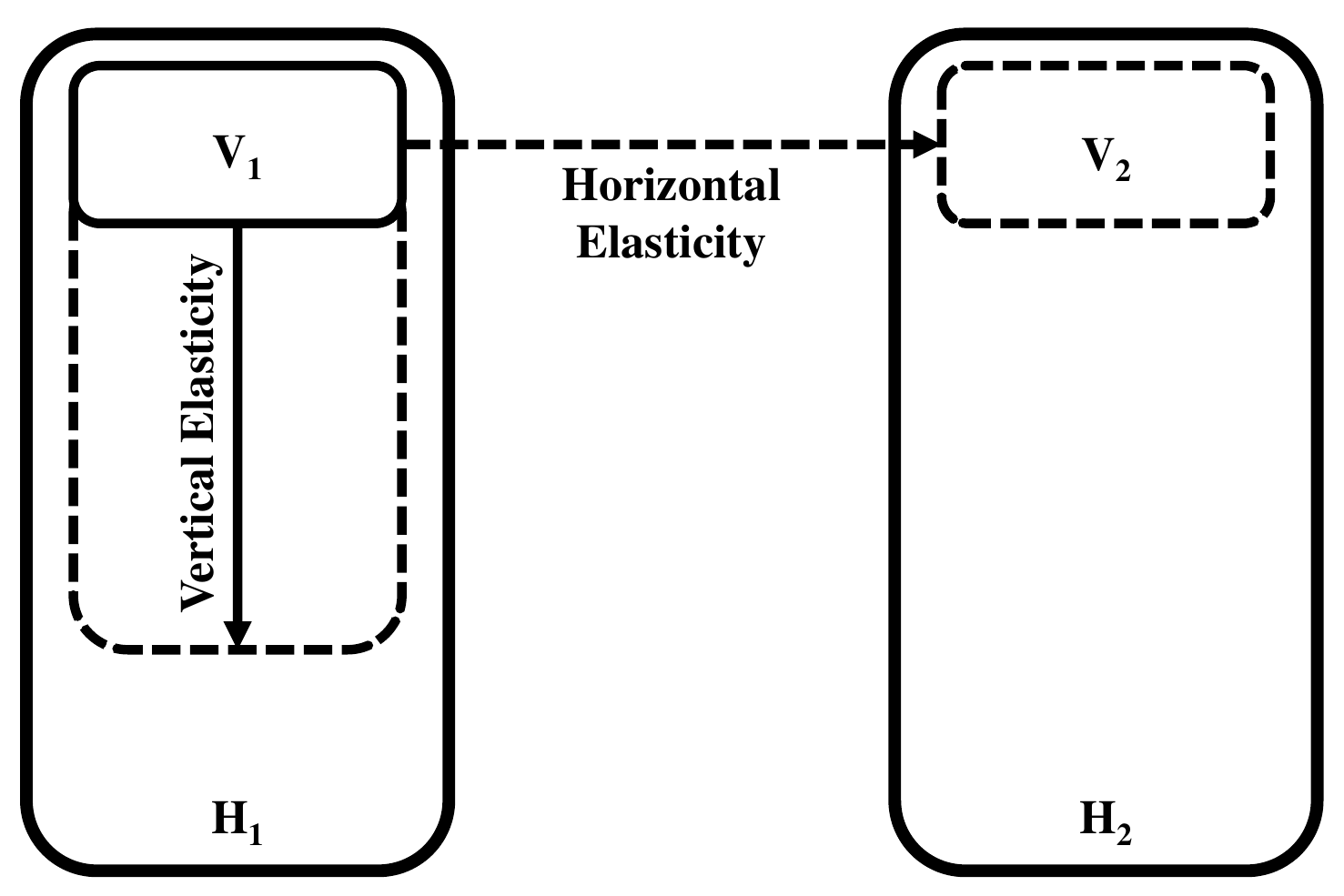}
\caption{Vertical and horizontal elasticity. Vertical elasticity dynamically adjusts the capacities of virtual resources inside a VM while horizontal elasticity dynamically adjusts the number of VMs (e.g. in distributed applications).}
\label{fig:elasticity}
\end{figure}

\subsection{Elasticity}
\label{sec:elasticity}
Considering the dynamic workload of modern cloud applications, proactive elasticity is a very important issue to address for CSPs in order to deal with under-provisioning (saturation) and over-provisioning (under-utilization) of cloud resources \cite{armbrust2009m}. Under-provisioning can cause SLA violations, impacting directly on economical revenue while over-provisioning can cause inefficient utilization of resources, directly impacting on resource utilization and energy consumption.

Research articles considering online formulations of the provider-oriented VMP literature have already studied two types of elasticity: vertical and horizontal (see Figure \ref{fig:elasticity}). Vertical elasticity can be defined as the ability of cloud services to dynamically change capacities of virtual resources (e.g. CPU and RAM memory) inside a VM, while horizontal elasticity can be defined as the ability of cloud services to dynamically adjust the number of VMs \cite{Wang2012}. 

It should be noted that from a CSP perspective, cloud services considering elasticity should be more important (i.e. higher level of SLA) than other non-elastic cloud services. An elastic cloud application could request additional resources to scale up the applications' resources and a CSP must consider more important the mentioned request than a request from a not elastic cloud service.   

Implementing vertical elasticity requires shorter time of service reconfiguration than horizontal elasticity, but with a higher migration cost. On the other hand, horizontal elasticity enables stronger high availability than vertical elasticity, but a coordination overhead is required and infrastructure complexity increases \cite{Wang2012}.

\subsection{Overbooking}
\label{sec:overbooking}
Resources overbooking can make cloud services more profitable for CSPs, overlaying requested virtual resources onto physical resources at a higher ratio than 1:1 \cite{hoeflin2012quantifying}. Online formulations of the provider-oriented VMP considering overbooking include particular considerations to efficiently attend customers' requirements, enforcing SLAs. 

Considering the dynamic workload of cloud applications and services, virtual resources of VMs are also dynamically used giving space to re-utilization of idle resources that were already reserved. Research articles considering online formulations of the provider-oriented VMP literature have already studied two types of overbooking: server and network resources overbooking. 

In this context, CSPs should measure the utilization of resources of VMs in order to correctly manage the overbooking with the available physical resources, minimizing SLA violations. Monitoring utilization of virtual resources and workload of cloud applications and services also helps CSPs to consider forecasting techniques for approximating in advance the required management actions (e.g. migrations of VMs) for the consolidation process, to reduce resource under-provisioning \cite{fang2013vmplanner,zhang2014dynamic}.

\section{Proposed Taxonomy}
\label{taxonomy}
Based on the universe of 64 studied research articles, dynamic environments for online formulations of the provider-oriented VMP problem may be classified by one or more of the following classification criteria: (1) elasticity and (2) overbooking, as presented in Section \ref{criteria}. 

The proposed taxonomy is presented in Figure \ref{table_taxonomy} as a two-dimensional coordinate axis where each dimension represents a classification criteria (elasticity and overbooking).

First, dynamic environments could be formulated considering one of the following elasticity values: 

\begin{itemize}	
\item elasticity=0: no elasticity;
\item elasticity=1: horizontal elasticity;
\item elasticity=2: vertical elasticity;
\item elasticity=3: horizontal and vertical elasticity. 
\end{itemize}

Additionally, identified dynamic environments may also consider one of the following overbooking values: 

\begin{itemize}
\item overbooking=0: no overbooking;
\item overbooking=1: server resources overbooking;
\item overbooking=2: network resources overbooking;
\item overbooking=3: server and network overbooking.
\end{itemize}

Based on the combinations of the possible values of the classification criteria (elasticity, overbooking), the proposed taxonomy identified 16 different possible environments (see Figure \ref{table_taxonomy}). Considering this representation, each identified dynamic environment is denoted by its elasticity and overbooking coordinates. For example, Environment (0,0) denotes a dynamic environment that does not consider neither any type of elasticity nor any type of overbooking, while Environment (1,3) denotes a dynamic environment that considers horizontal elasticity with both types of overbooking. 

It should be mentioned that all the identified environments in this work consider that VMs are dynamically created and destroyed. According to the identified dynamic environments, the simplest environment is Environment (0,0), while the most complex environment is Environment (3,3). Additionally, the proposed taxonomy showed that 50\% of the articles studied Environment (0,1) while 39\% of the studied articles considered Environment (0,0), representing the most studied environments in the considered literature. Several unexplored environments were also identified, as detailed in the following subsections.

\subsection{Cloud Service and Environment Notation}
\label{cs_notation}
CSPs dynamically receive requests for the placement of cloud services with different characteristics according to the classification criteria presented in Section \ref{criteria}, representing real-world environments and generalizing the deployment of cloud services in several possible cloud architectures (e.g. single-cloud, distributed-cloud or federated-cloud). Cloud services may represent simple services such as Domain Name Service (DNS) or complex multi-tier elastic applications.

A cloud service is composed by a set of VMs, where each VM of a cloud service could be located for its execution in different cloud datacenters according to the customers preferences or requirements (e.g. legal issues or high-availability). 

Configuration of VMs of a cloud service changes dynamically when elasticity is considered. On the other hand, utilization of virtual resources change dynamically according to the demand when overbooking is considered; otherwise, the utilization of each virtual resource is considered at 100\%.

Formally, a cloud service $S_{b}$ can be distributed across different possible cloud datacenters. Each cloud datacenter $DC_c$ hosts VMs $V'_{cj}$ associated to different cloud services. A VM $V'_{cj}$ associated to a service $S_{b}$ is denoted as $V''_{bcj}$.
\\\\
\noindent where:

\begin{tabbing}
\hspace*{1.65cm}\=\kill
	$S_{b}$:	\>Cloud service $b$;\\
	$DC_{c}$: \>Cloud datacenter $c$;\\
	$mDC_{c}$: \>Number of VMs $V_{j}$ in $DC_{c}$;\\	
	$mS_{b}$: \>Number of VMs $V_{j}$ in $S_{b}$;\\	
	$V'_{cj}$:	\>$V_j$ in $DC_{c}$;\\
	$V''_{bcj}$:\>$V_j$ in $DC_{c}$ from service $S_{b}$.\\
\end{tabbing}

Figure \ref{fig:service} presents a basic example of a cloud service $S_1$, distributed across 2 cloud datacenters $DC_1$ and $DC_2$ and using 4 VMs: $V''_{111}$, $V''_{112}$, $V''_{121}$, $V''_{122}$. These cloud datacenters could represent geo-distributed datacenters owned by one CSP or a federated-cloud with two different CSPs. Each cloud datacenter hosts 2 VMs of $S_1$: $V'_{11}$ and $V'_{12}$ represent $V_{1}$ and $V_{2}$ in $DC_1$ respectively (analogously $DC_2$ hosts 2 VMs).

\begin{figure}[b]
\centering
\includegraphics[scale=0.5]{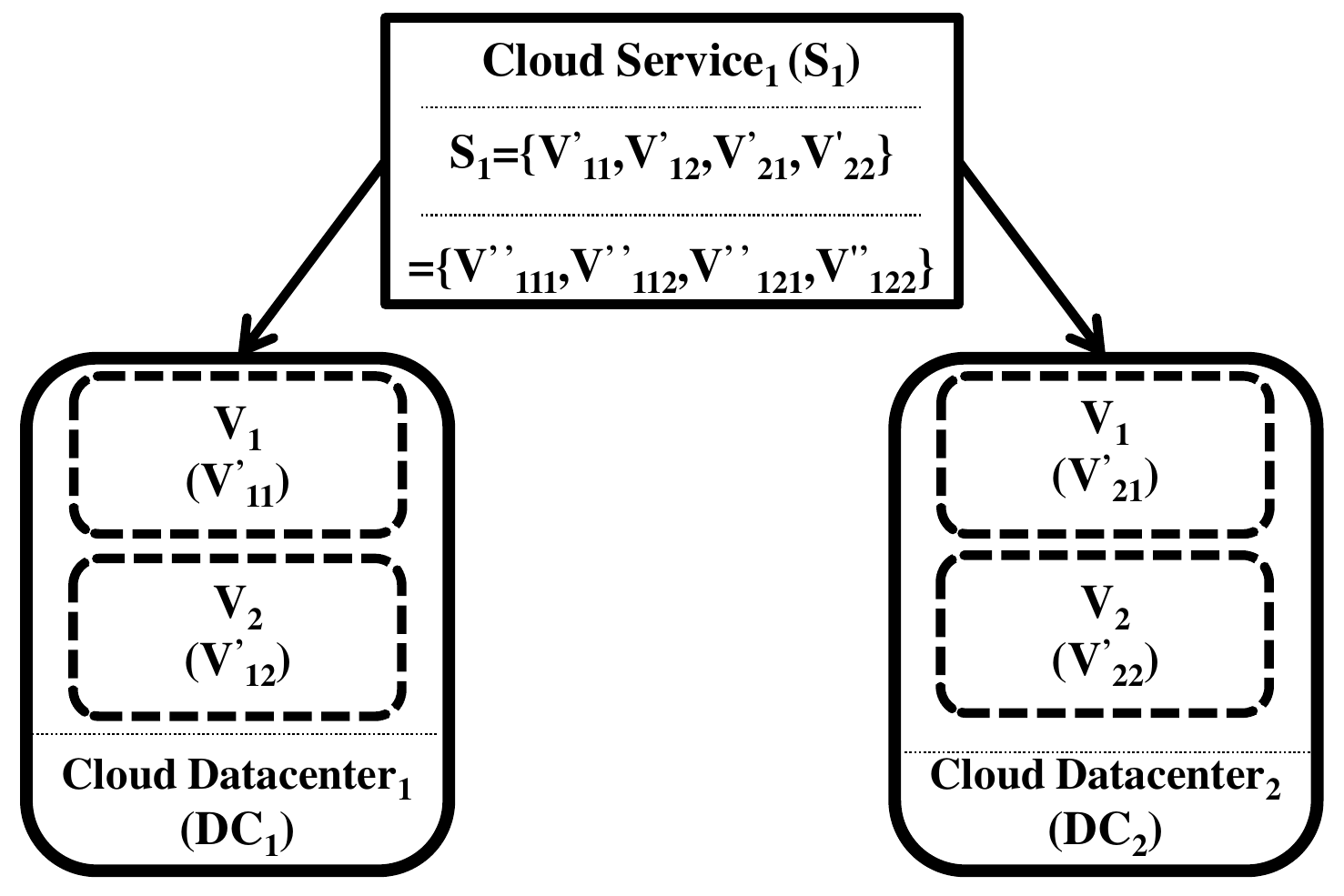}
\caption{Example of a cloud service considered in this work.}
\label{fig:service}
\end{figure}

Complementing the above notation, each cloud datacenter $DC_c$ may be represented as:

\begin{equation}
\label{eq_1}  
DC_{c}=\{V'_{cj}\}, j \in \{1,2, \dots, mDC_c\}
\end{equation}

For simplicity, this work considers only processing, memory and network resources for a VM, but the notation is general enough for considering any set of virtual resources.

\begin{equation}
\label{eq_2}  
\begin{aligned}
V''_{bcj}=&\{Vcpu''_{bcj},Vram''_{bcj},Vnet''_{bcj},R''_{bcj},\\
&SLA''_{bcj},t_{init},t_{end}\}
\end{aligned}
\end{equation}

\noindent where:

\begin{tabbing}
\hspace*{1.65cm}\=\kill
	$V''_{bcj}$:		\>$V_j$ in $DC_{c}$ from service $S_{b}$;\\
	$Vcpu''_{bcj}$:	\>Processing requirements of $V''_{bcj}$ in [ECU];\\
	$Vram''_{bcj}$:	\>Memory requirements of $V''_{bcj}$ in [GB];\\	
	$Vnet''_{bcj}$:	\>Network requirements of $V''_{bcj}$ in [Mbps];\\
	$R''_{bcj}$:		\>Economical revenue for locating $V''_{bcj}$ in [\$];\\
	$SLA''_{bcj}$:	\>SLA of $V''_{bcj}$. $SLA''_{bcj}$ $\in$ $\{1,..,s\}$;\\
	$s$: 						\>Highest priority level of SLAs;\\
	$t_{init}$: 		\>Initial discrete time when $V''_{bcj}$ is executed;\\
	$t_{end}$:			\>Final discrete time when $V''_{bcj}$ is executed.\\
\end{tabbing}
	
Utilization of the resources of each $V''_{bcj}$ is represented by:

\begin{equation}
\label{eq_3}  
U''_{bcj}=\{Ucpu''_{bcj},Uram''_{bcj},Unet''_{bcj}\}
\end{equation}
	
\noindent where:

\begin{tabbing}
\hspace*{1.65cm}\=\kill
	$U''_{bcj}$:		\>Utilization of requirement $V''_{bcj}$;\\
	$Ucpu''_{bcj}$:	\>Utilization of $Vcpu''_{bcj}$ in [ECU]; \\
	$Uram''_{bcj}$:	\>Utilization of $Vram''_{bcj}$ in [GB];\\	
	$Unet''_{bcj}$:	\>Utilization of $Vnet''_{bcj}$ in [Mbps].\\
\end{tabbing}

Note that in practical applications $Ucpu''_{bcj}$ is lower than $Vcpu''_{bcj}$ ($Ucpu''_{bcj} \leq Vcpu''_{bcj}$), giving place to CPU overbooking of resources. The same overbooking situation may occur for other resources.

Each of the 16 identified environments (see Figure \ref{table_taxonomy}) considers different parameters that dynamically change as a  function of time $t$, giving place to possible different notations for each environment. The following subsections detail each of the identified dynamic environment, presenting particular notation for its characterization.


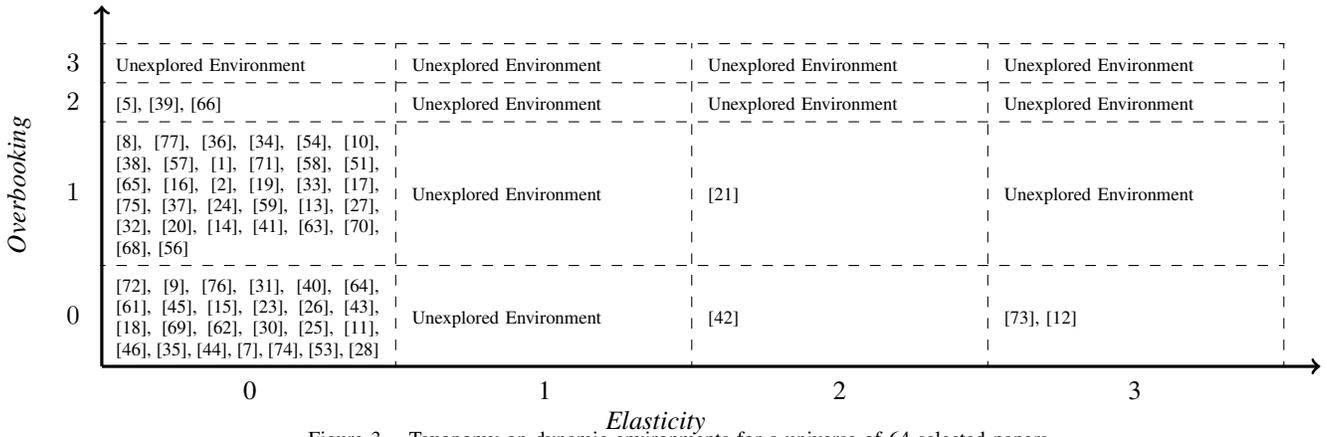
\begin{figure*}[t]

\centering
\renewcommand{\arraystretch}{1.2}

\begin{tikzpicture}[overlay]
\node[text width=2cm,rotate=90] at (-8.8,-1.8) {\textit{Overbooking}};
\draw[very thick,->] (-7.7,-4.3) -- (-7.7, 0.5);
\end{tikzpicture}

\begin{tabular}{m{0.3cm}m{3.5cm}:m{3.5cm}:m{3.5cm}:m{3.5cm}:m{0.0cm}}

\cdashline{2-5}
\centering
\bfseries $3$ & 
\scriptsize Unexplored Environment &
\scriptsize Unexplored Environment &
\scriptsize Unexplored Environment & 
\scriptsize Unexplored Environment & 
\\

\cdashline{2-5}
\centering
\bfseries $2$ &
\scriptsize \cite{Biran2012,Kantarci2012,song2014optimization} &
\scriptsize Unexplored Environment &
\scriptsize Unexplored Environment & 
\scriptsize Unexplored Environment & 
\\ 

\cdashline{2-5}
\centering
\bfseries $1$ &
\scriptsize \cite{cao2014energy,zhang2014dynamic,Jin2013,hwang2013hierarchical,moreno2013improved,chen2013intelligent,kakadia2013network,prevost2013optimal,alicherry2013optimizing,wang2013particle,ribas2013pbfvmc,lu2013qos,singh2013reduce,dong2013virtual,anand2013virtual,fang2013vmplanner,Huang2012,Dupont2012,Zamanifar2012,Jin2012,Goudarzi2012,Ribas2012,Dias2012,Gupta2012,Huang2012b,Ferreto2011,Do2011,Le2011,Shi2011,Tsakalozos2011,Speitkamp2010,Piao2010} & 
\scriptsize Unexplored Environment & 
\scriptsize \cite{Ferreto2011a} & 
\scriptsize Unexplored Environment & 
\\

\cdashline{2-5}
\centering
\bfseries $0$ &
\scriptsize \cite{wang2014eqvmp,Chang2013,Zhang2013,Huang2013,Kord2013,Shigeta2013,Sato2013,li2013energy,Dong2013,georgiou2013exploiting,gupta2013hpc,li2013migration,fang2013power,tsai2013prevent,shi2013provisioning,hong2013qoe,guo2013shadow,dalvandi2013time,Lin2012,Jiang2012a,Li2012a,Calcavecchia2012,Wu2012b,Mishra2011,Ho2011} & 
\scriptsize Unexplored Environment &
\scriptsize \cite{Li2013a} & 
\scriptsize \cite{Wang2012,dang2013higher} & 
\\
\end{tabular}

\begin{tikzpicture}[overlay]
\draw[very thick,->] (-7.7,0) -- (8.5, 0);
\node[text width=2cm] at (0,-0.75) {\textit{Elasticity}};
\end{tikzpicture}

\begin{tabular}{m{0.3cm}m{3.5cm}m{3.5cm}m{3.5cm}m{3.5cm}m{0.0cm}}
\centering

\bfseries & \centering 0 & \centering 1 & \centering 2 & \centering 3 & \hfill \\ 
\end{tabular}
\caption{Taxonomy on dynamic environments for a universe of 64 selected papers.}
\label{table_taxonomy}
\end{figure*}

\subsection{Dynamic Environment Classification}
\label{environments}
Based on the notation previously presented in Section \ref{cs_notation}, the 16 identified environments (see Figure \ref{table_taxonomy}) present particular considerations and different time variables may be defined for a correct characterization. A summary of the time variables is presented in Table \ref{table_dynamic_notation}. Next, the identified dynamic environments are presented, enumerated by its elasticity and overbooking coordinates.

\begin{itemize}
\item \textbf{(0,0)} \textit{No Elasticity (0), No Overbooking (0)}: It represents the most basic dynamic environment identified for solving the provider-oriented VMP problem. The CSP have to attend the placement of cloud services that are dynamically created and destroyed in function of time $t$. From the studied universe of 64 articles, 39\% proposed formulations of the VMP problem considering this basic environment (Figure \ref{table_taxonomy}). 

\item \textbf{(0,1)} \textit{No Elasticity (0), Server Resources Overbooking (1)}: According to the studied articles, the provider-oriented VMP problem is mostly formulated considering overbooking of server resources (i.e. processing, memory and storage) without considering neither horizontal nor vertical elasticity. This environment represents 50\% of the studied universe of 64 articles (see Figure \ref{table_taxonomy}). For this particular environment, CSPs must monitor the dynamic utilization of virtual server resources for a safe overbooking. The following variables must be defined as a function of time: $Ucpu''_{bcj}(t)$ and $Uram''_{bcj}(t)$.

\item \textbf{(0,2)} \textit{No Elasticity (0), Network Resources Overbooking (2)}: Overbooking could be also considered exclusively for virtual network resources. Analogously to the Environment (0,1), CSPs must monitor the dynamic utilization of virtual network resources for a safe overbooking. Consequently, the utilization of network resources is defined as a time variable: $Unet''_{bcj}(t)$. This environment represents only 5\% of the studied articles (see Figure \ref{table_taxonomy}).

\item \textbf{(0,3)} \textit{No Elasticity (0), Server and Network Resources Overbooking (3)}: Representing the most complex environment for overbooking, this dynamic environment is identified as a research opportunity, considering that no studied article proposed a formulation of the provider-oriented VMP problem that jointly considers overbooking of server and network resources without elasticity (see Figure \ref{table_taxonomy}). For this particular environment, CSPs must monitor the dynamic utilization of both virtual server and network resources for a safe overbooking. Consequently, $Ucpu''_{bcj}(t)$, $Uram''_{bcj}(t)$ and $Unet''_{bcj}(t)$ are defined as a function of time.

\item \textbf{(1,0)} \textit{Horizontal Elasticity (1), No Overbooking (0)}: Elasticity could be considered in order to efficiently attend the dynamic demand of resources according to a SLA associated to a given cloud service. A dynamic environment that considers horizontal elasticity represents particular considerations associated to scaling up and down the number of requested VMs that composes a cloud service. 

Determining when to scale is not considered a responsibility of the CSPs and it is out of the scope of this work. In this context, the number of VMs of cloud services is a time variable: $mS_{b_{min}} \leq mS_b(t) \leq mS_{b_{max}}$. This particular environment is also identified as a research opportunity, representing the most basic environment for solving problems considering horizontal elasticity for parallel applications such as MapReduce jobs.

\item \textbf{(1,1)} \textit{Horizontal Elasticity (1), Server Resources Overbooking (1)}: Additionally to horizontal elasticity, a dynamic environment could also include overbooking of server resources. The following variables are defined as a function of time: $Ucpu''_{bcj}(t)$, $Uram''_{bcj}(t)$ and $mS_b(t)$. This environment also represents a research opportunity.

\item \textbf{(1,2)} \textit{Horizontal Elasticity (1), Network Resources Overbooking (2)}: Analogously to the Environment (1,1), CSPs must monitor the dynamic utilization of virtual network resources for a safe overbooking. Consequently, $Unet''_{bcj}(t)$ is defined as a time variable, additionally to $mS_b(t)$. This environment is identified as a research opportunity, considering that no studied article considers overbooking of network resources with horizontal elasticity.

\item \textbf{(1,3)} \textit{Horizontal Elasticity (1), Server and Network Resources Overbooking (3)}: No studied article proposed a formulation of the provider-oriented VMP problem that jointly considers overbooking of server and network resources with horizontal elasticity, representing a research opportunity. For this particular environment, the following variables are defined as a function of time: $Ucpu''_{bcj}(t)$, $Uram''_{bcj}(t)$, $Unet''_{bcj}(t)$ and $mS_b(t)$.

\item \textbf{(2,0)} \textit{Vertical Elasticity (2), No Overbooking (0)}: As mentioned before, elasticity could be considered in order to efficiently attend the dynamic demand of resources according to a SLA associated to a cloud service. A dynamic environment that considers vertical elasticity represents particular considerations associated to the virtual resources capacities of requested VMs that composes a cloud service. This work considers processing and memory requirements as time variables: $Vcpu''_{bcj}(t)$ and $Vram''_{bcj}(t)$. 

It should be mentioned that the notation presented in this section is general enough to consider any other resources for vertical elasticity such as $Vnet''_{bcj}(t)$ just to cite one. According to the studied articles, only \cite{Li2013a} (1.5\%) studied this environment (see Figure \ref{table_taxonomy}).

\item \textbf{(2,1)} \textit{Vertical Elasticity (2), Server Resources Overbooking (1)}: Additionally to vertical elasticity, a dynamic environment could also include overbooking of server resources. For these particular environment, CSPs must monitor the dynamic utilization of virtual server resources for a safe overbooking. Consequently, the following variables are defined as a function of time: $Ucpu''_{bcj}(t)$, $Uram''_{bcj}(t)$, $Vcpu''_{bcj}(t)$ and $Vram''_{bcj}(t)$. According to the studied articles, only \cite{Ferreto2011a} (1.5\%) studied this environment (see Figure \ref{table_taxonomy}).

\item \textbf{(2,2)} \textit{Vertical Elasticity (2), Network Resources Overbooking (2)}: Analogously to the Environment (2,1), $Unet''_{bcj}(t)$ is defined as a time variable, additionally to $Vcpu''_{bcj}(t)$ and $Vram''_{bcj}(t)$. This particular environment represents a research opportunity, considering that no studied article proposed a formulation of the provider-oriented VMP in this particular environment.

\item \textbf{(2,3)} \textit{Vertical Elasticity (2), Server and Network Resources Overbooking (3)}: No studied article proposed a formulation of the provider-oriented VMP problem that jointly considers overbooking of server and network resources with vertical elasticity. For this particular environment, the following variables are defined as a function of time: $Ucpu''_{bcj}(t)$, $Uram''_{bcj}(t)$, $Unet''_{bcj}(t)$, $Vcpu''_{bcj}(t)$ and $Vram''_{bcj}(t)$.

\item \textbf{(3,0)} \textit{Horizontal and Vertical Elasticity (3), No Overbooking (0)}: Both types of elasticity lead to different impacts for cloud datacenters infrastructure management and respond to different requirements of customers' applications. Definitely, in real world environments CSPs should be able to solve the VMP problem considering formulations that jointly implement both horizontal and vertical elasticity for cloud services. In this context, \cite{Wang2012} and \cite{dang2013higher} proposed different approaches for dealing with these challenges, representing 3\% of the studied universe (see Figure \ref{table_taxonomy}). 

In this environment of both mixed elasticity types, the following time variables are defined: $mS_b(t)$, $Vcpu''_{bcj}(t)$ and $Vram''_{bcj}(t)$.

\item \textbf{(3,1)} \textit{Horizontal and Vertical Elasticity (3), Server Resources Overbooking (1)}: Additionally to horizontal and vertical elasticity, a dynamic environment could also include overbooking of server resources. For these particular environment, CSPs must monitor the dynamic utilization of virtual server resources for a safe overbooking. Consequently, the following variables are defined as a function of time: $Ucpu''_{bcj}(t)$, $Uram''_{bcj}(t)$, $mS_b(t)$, $Vcpu''_{bcj}(t)$ and $Vram''_{bcj}(t)$.

\item \textbf{(3,2)} \textit{Horizontal and Vertical Elasticity (3), Network Resources Overbooking (2)}: Analogously to the Environment (3,1), the following variables are defined in function of time: $mS_b(t)$, $Vcpu''_{bcj}(t)$, $Vram''_{bcj}(t)$ and $Unet''_{bcj}(t)$. This particular environment represent a research opportunity, considering that no studied article proposed a formulation of the provider-oriented VMP in this particular environment.

\item \textbf{(3,3)} \textit{Horizontal and Vertical Elasticity (3), Server and Network Resources Overbooking (3)}: Considering both types of elasticity and both types of overbooking represent the most complex environment identified in this work. CSPs efficiently solving formulations of the VMP problem in this complex dynamic environment will represent a considerable advance on this research area and its cloud datacenters will be able to scale according to trending types of requirements with sufficient flexibility. As the most general environment, the following variables are defined in function of time for characterizing this complex environment: $mS_b(t)$, $Vcpu''_{bcj}(t)$, $Vram''_{bcj}(t)$, $Ucpu''_{bcj}(t)$, $Uram''_{bcj}(t)$ and $Unet''_{bcj}(t)$. A recommended path for research is exploring and addressing challenges of particular environments identified as research opportunities before considering this advanced and complete dynamic environment for solving the provider-oriented VMP problem.
\end{itemize}

\begin{table}[t]
\caption{Summary of time variables for the 16 identified dynamic environments.}
\label{table_dynamic_notation}
\begin{tabular}{|l|l|l|m{1.9cm}|}
\hline
\bfseries Env. & \bfseries Elasticity Type & \bfseries Overbooking Type & \bfseries Time Variables\\
\hline
(0,0) & Not Considered & Not Considered & - \\ 
\hline
(0,1) & Not Considered & Server & -$Ucpu''_{bcj}(t)$ -$Uram''_{bcj}(t)$ \\
\hline
(0,2) & Not Considered & Network & -$Unet''_{bcj}(t)$ \\
\hline
(0,3) & Not Considered & Server and Network & -$Ucpu''_{bcj}(t)$ -$Uram''_{bcj}(t)$ -$Unet''_{bcj}(t)$\\ 
\hline
(1,0) & Horizontal & Not Considered & -$mS_b(t)$ \\
\hline
(1,1) & Horizontal & Server & -$mS_b(t)$ \hspace{1cm} -$Ucpu''_{bcj}(t)$ -$Uram''_{bcj}(t)$ \\
\hline
(1,2) & Horizontal & Network & -$mS_b(t)$ \hspace{1cm} -$Unet''_{bcj}(t)$ \\
\hline
(1,3) & Horizontal & Server and Network & -$mS_b(t)$ \hspace{1cm} -$Ucpu''_{bcj}(t)$ -$Uram''_{bcj}(t)$ -$Unet''_{bcj}(t)$ \\ 
\hline
(2,0) & Vertical & Not Considered & -$Vcpu''_{bcj}(t)$ -$Vram''_{bcj}(t)$ \\
\hline
(2,1) & Vertical & Server & -$Vcpu''_{bcj}(t)$ -$Vram''_{bcj}(t)$ -$Ucpu''_{bcj}(t)$ -$Uram''_{bcj}(t)$ \\
\hline
(2,2) & Vertical & Network & -$Vcpu''_{bcj}(t)$ -$Vram''_{bcj}(t)$ -$Unet''_{bcj}(t)$ \\
\hline
(2,3) & Vertical & Server and Network & -$Vcpu''_{bcj}(t)$ -$Vram''_{bcj}(t)$ -$Ucpu''_{bcj}(t)$ -$Uram''_{bcj}(t)$ -$Unet''_{bcj}(t)$ \\ 
\hline
(3,0) & Horizontal and Vertical & Not Considered & -$mS_b(t)$ \hspace{1cm} -$Vcpu''_{bcj}(t)$ -$Vram''_{bcj}(t)$ \\
\hline
(3,1) & Horizontal and Vertical & Server & -$mS_b(t)$ \hspace{1cm} -$Vcpu''_{bcj}(t)$ -$Vram''_{bcj}(t)$ -$Ucpu''_{bcj}(t)$ -$Uram''_{bcj}(t)$  \\
\hline
(3,2) & Horizontal and Vertical & Network & -$mS_b(t)$ \hspace{1cm} -$Vcpu''_{bcj}(t)$ -$Vram''_{bcj}(t)$ -$Unet''_{bcj}(t)$ \\
\hline
(3,3) & Horizontal and Vertical & Server and Network & -$mS_b(t)$ \hspace{1cm} -$Vcpu''_{bcj}(t)$ -$Vram''_{bcj}(t)$ -$Ucpu''_{bcj}(t)$ -$Uram''_{bcj}(t)$ -$Unet''_{bcj}(t)$ \\ 
\hline
\end{tabular}
\end{table}

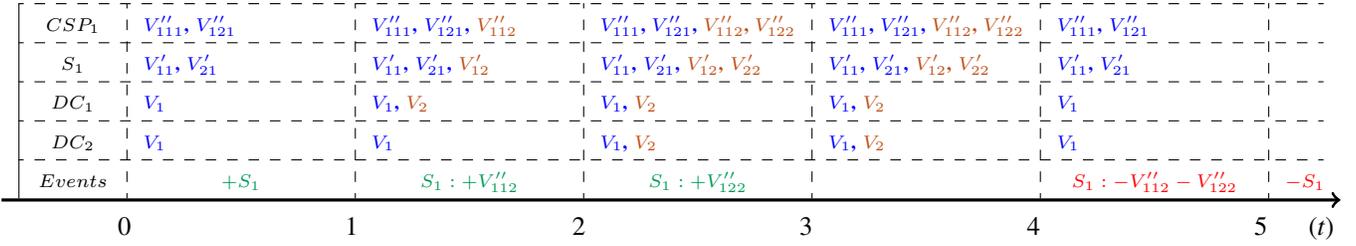
\begin{figure*}[t]
\centering
\renewcommand{\arraystretch}{1.2}
\begin{tabular}{|m{1.0cm}:m{2.6cm}:m{2.6cm}:m{2.6cm}:m{2.6cm}:m{2.6cm}:m{0.3cm}}
\cdashline{1-7}
\centering
\scriptsize  \bfseries $CSP_1$ &
\scriptsize  \bfseries \color{blue} $V''_{111}$, $V''_{121}$ & 
\scriptsize  \bfseries \color{blue} $V''_{111}$, $V''_{121}$, \color{Bittersweet} $V''_{112}$ & 
\scriptsize  \bfseries \color{blue} $V''_{111}$, $V''_{121}$, \color{Bittersweet} $V''_{112}$, $V''_{122}$ &
\scriptsize  \bfseries \color{blue} $V''_{111}$, $V''_{121}$, \color{Bittersweet} $V''_{112}$, $V''_{122}$ &
\scriptsize  \bfseries \color{blue} $V''_{111}$, $V''_{121}$ & \
\\

\cdashline{1-7}
\centering
\scriptsize  \bfseries $S_1$ &
\scriptsize  \bfseries \color{blue} $V'_{11}$, $V'_{21}$ & 
\scriptsize  \bfseries \color{blue} $V'_{11}$, $V'_{21}$, \color{Bittersweet} $V'_{12}$ & 
\scriptsize  \bfseries \color{blue} $V'_{11}$, $V'_{21}$, \color{Bittersweet} $V'_{12}$, $V'_{22}$ & 
\scriptsize  \bfseries \color{blue} $V'_{11}$, $V'_{21}$, \color{Bittersweet} $V'_{12}$, $V'_{22}$ & 
\scriptsize  \bfseries \color{blue} $V'_{11}$, $V'_{21}$ \
\\

\cdashline{1-7}
\centering
\scriptsize \bfseries $DC_1$ &
\scriptsize \bfseries \color{blue} $V_{1}$ &  
\scriptsize \bfseries \color{blue} $V_{1}$, \color{Bittersweet} $V_{2}$ &  
\scriptsize \bfseries \color{blue} $V_{1}$, \color{Bittersweet} $V_{2}$ &  
\scriptsize \bfseries \color{blue} $V_{1}$, \color{Bittersweet} $V_{2}$ &  
\scriptsize \bfseries \color{blue} $V_{1}$ \
\\ 

\cdashline{1-7}
\centering
\scriptsize \bfseries $DC_2$ &
\scriptsize \bfseries \color{blue} $V_{1}$ &  
\scriptsize \bfseries \color{blue} $V_{1}$ &  
\scriptsize \bfseries \color{blue} $V_{1}$, \color{Bittersweet} $V_{2}$ &  
\scriptsize \bfseries \color{blue} $V_{1}$, \color{Bittersweet} $V_{2}$ &  
\scriptsize \bfseries \color{blue} $V_{1}$ \
\\

\cdashline{1-7}
\centering
\scriptsize \bfseries $Events$ &
\scriptsize \bfseries \centering \color{ForestGreen} $+S_1$ &
\scriptsize \bfseries \centering \color{ForestGreen} $S_1: +V''_{112}$ &
\scriptsize \bfseries \centering \color{ForestGreen} $S_1: +V''_{122}$ &
\scriptsize \bfseries $ $ &
\scriptsize \bfseries \centering \color{red} $S_1: -V''_{112} -V''_{122}$ &
\scriptsize \bfseries \color{red} $-S_1$ \
\\
\end{tabular}

\begin{tikzpicture}
\draw[very thick,->] (0,0) -- (17.8,0);
\end{tikzpicture}

\begin{tabular}{m{1.0cm}m{2.6cm}m{2.6cm}m{2.6cm}m{2.6cm}m{2.6cm}m{0.3cm}m{0.3cm}}
\centering
\bfseries & 0 & 1 & 2 & 3 & 4 & 5 & (\textit{t})\\ 
\end{tabular}
\caption{Basic example of Environment (1,0)}
\label{fig:environment_16}
\end{figure*}

\subsection{Dynamic Environment Examples}
Due to space limitations, this work focus on a representative example of the Environment (1,0), representing an elastic application implementing horizontal elasticity (see Figure \ref{fig:environment_16}). Interested readers can refer to \cite{ortigoza2015} for a complete set of examples of the 16 identified dynamic environments.

It is important to remember that Environment (1,0) includes only horizontal elasticity, but vertical elasticity as well as both types of overbooking (server and network resources) can be observed in the workload trace presented in Section \ref{workload_example}. 

Figure \ref{fig:environment_16} presents different levels of detail for the environment. The CSP level ($CSP_1$) represents the requests that CSPs receive for the placement of cloud services (or VMs) in the PMs of the available cloud datacenters. Next, cloud service level ($S_1$) details requested resources of cloud services at each discrete time. Cloud datacenter levels ($DC_1$ and $DC_2$) detail resources of cloud services for each cloud datacenter.

The horizontal elasticity of the example in Figure \ref{fig:environment_16} can be observed considering that initially, cloud service $S_1$ starts in $t=0$ requesting 2 VMs (in blue) across cloud datacenters $DC_1$ and $DC_2$ from $t=0$ to $t=5$. Assuming an increasing demand for resources, $S_1$ scales up the number of VMs adding 1 VM hosted at cloud datacenter $DC_1$ (in brown) at $t=1$. In $t=2$, $S_1$ scales up the number of VMs adding 1 more VM hosted at cloud datacenter $DC_2$ (in brown) resulting in 4 VMs for attending the demand for resources from $t=2$ to $t=3$. The cloud service $S_1$ returns to its initial configuration scaling down to 2 VMs at $t=4$, assuming a decreasing demand for resources, finishing the requests for resources at $t=5$.

\section{Workload Traces for Dynamic Environments}
\label{workloads}
Based on the proposed taxonomy, several research opportunities have been identified (see Section \ref{environments}). Therefore, sample workload traces for each particular environment are required in order to: (1) explore the challenges associated to each environment, (2) propose formulations and test algorithms that solve these challenges with different workload types and (3) effectively compare performance and quality of different algorithms with reproducible experiments.

As identified in \cite{lopez2014survey}, there is no existing testbed problem instances for the VMP that can today be used as a world accepted benchmark. Consequently, the authors are working on a workload trace generator for the VMP problem to be able to generate different instances for experimental tests based on the dynamic environments proposed in this work. A brief introduction of preliminary results is presented in this section.

\begin{table}[b]
\centering
\caption{Input data for example workload trace from Table \ref{table_workload}.}
\label{table_inputs}
\begin{tabular}{|l|c|}
\hline
\bfseries Input Data & \bfseries Value (Min - Max) \\
\hline
Workload trace duration ($t$) & (4 - 4) \\ 
\hline
Range of CPU values for resources & (4 - 10) \\
\hline
Range of Memory values for resources & (2 - 16) \\
\hline
Range of Network values for resources & (100 - 1000) \\
\hline
Range of CPU values for utilization & (2 - 10) \\
\hline
Range of Memory values for utilization & (1 - 16) \\
\hline
Range of Network values for utilization & (0 - 1000) \\
\hline
Range of revenue values for executing VMs & (0.1 - 1.5) \\ 
\hline
Range of SLA values of VMs & (0 - 2) \\
\hline
Range of number of VMs for cloud services & (1 - 6) \\
\hline
Number of cloud services & (1 - 1) \\
\hline
Probability distribution & Random \\
\hline
\end{tabular}
\end{table}

\begin{table*}[t]
\centering
\caption{Example of workload trace for VMP problem in Environment (3,3)}
\label{table_workload}
\begin{tabular}{|c|c|c|c|c|c|c|c|c|c|c|c|}
\hline
\bfseries $t$ & \bfseries $S_b$ & \bfseries $D_c$ & \bfseries $V_j$ & \bfseries $Vcpu''_{bcj}$ & \bfseries $Vram''_{bcj}$ & \bfseries $Vnet''_{bcj}$ & \bfseries $R''_{bcj}$ & \bfseries $SLA''_{bcj}$ & \bfseries $Ucpu''_{bcj}$ & \bfseries $Uram''_{bcj}$ & \bfseries $Unet''_{bcj}$\\
\hline
0 &	1 &	1 &	1 &	8 &	16 & 	1000 &	0.5 &	1 &	8 &	14 &	150\\
\hline
0 &	1 & 	2 &	1 &	8 &	16 &	1000 &	0.5 &	1 &	8 &	9 &	50\\
\hline
\hline											
1 &	1 &	1 &	1 &	8 &	16 & 	1000 &	0.5 &	1 &	7 &	10 &	160\\
\hline
1 &	1 & 	2 &	1 &	8 &	16 &	1000 &	0.5 &	1 &	7 &	10 &	100\\
\hline
1 &	1 & 	1 &	2 &	8 &	16 &	1000 &	0.5 &	1 &	7 &	7 &	70\\
\hline
\hline											
2 &	1 &	1 &	1 &	8 &	16 & 	1000 &	0.5 &	1 &	6 &	11 &	200\\
\hline
2 &	1 &	2 &	1 &	8 &	16 & 	1000 &	0.5 &	1 &	6 &	11 &	150\\
\hline
2 &	1 &	1 &	2 &	8 &	16 & 	1000 &	0.5 &	1 &	6 &	9 &	50\\
\hline
2 &	1 & 	2 &	2 &	8 &	16 &	1000 &	0.5 &	1 &	6 &	12 &	60\\
\hline
\hline 											
3 &	1 &	1 &	1 &	8 &	16 & 	1000 &	0.5 &	1 &	4 &	12 &	180\\
\hline
3 &	1 &	2 &	1 &	8 &	16 & 	1000 &	0.5 &	1 &	4 &	12 &	150\\
\hline
3 &	1 &	1 &	2 &	8 &	16 & 	1000 &	0.5 &	1 &	1 &	9 &	60\\
\hline
3 &	1 & 	2 &	2 &	8 &	16 &	1000 &	0.5 &	1 &	1 &	8 &	60\\
\hline
\hline 											
4 &	1 &	1 &	1 &	4 &	8 & 1000 &	0.5 &	1 &	2 &	6 &	200\\
\hline
4 &	1 & 2 &	1 &	4 &	8 &	1000 &	0.5 &	1 &	2 &	6 &	100\\
\hline
\end{tabular}
\end{table*}

The proposed workload trace generator for the VMP problem considers the following input data (see Table \ref{table_inputs}): 

\begin{itemize}
\item workload trace duration,
\item range of values for virtual resources of VMs,
\item range of values for utilization of virtual resources of VMs,
\item range of revenue values for executing VMs,
\item range of SLA values of VMs,
\item range of number of VMs for cloud services,
\item number of cloud services,
\item probability distribution.
\end{itemize}

It is important to mention that additionally, an user of the generator is also able to include real-world workload traces, extending or reducing the trace to specific requirements for the experiments. In this case, the workload generator could generate synthetic workloads based on real-world workloads adjusting the workload to a specific size of the problem instance (e.g. workload trace duration), mantaining characteristics of the real-world trace (e.g. probability distribution). 

\subsection{Workload Trace Generation Example}
\label{workload_example}
As an example of utilization of the workload trace generator, Table \ref{table_inputs} presents the input data considered for the generation of the workload trace presented in Table \ref{table_workload}. Due to space limitation as well as similarity of the structure of the workload traces of the different environments, Table \ref{table_workload} represents only a basic example of the most complex dynamic environment identified in this work, Environment (3,3). This example includes all possible dynamic parameters (resource capacities of VMs, number of VMs of a cloud service and utilization of resources of VMs). Interested readers can refer to \cite{ortigoza2015} for more detailed examples of workload traces for the 16 different dynamic environments.

According to Table \ref{table_inputs}, all the considered values are selected randomly from the specified values considering that for this particular example the probability distribution is uniform. It is important to mention that several other data probability distributions could be considered.

The generated workload trace of Table \ref{table_workload} could have only a duration of $t=4$ considering that both minimum and maximum values were 4 (see Table \ref{table_inputs}). For this particular example, the number of cloud services was fixed to 1 considering that both minimum and maximum values were 1 (see Table \ref{table_inputs}).

Additionally, the number of VMs of the cloud services in the example of Table \ref{table_workload} can be adjusted from 1 to 6. The revenue for executing VMs can vary from 0.1\$ to 1.5\$ and its corresponding SLAs can vary from 0 to 2. Analogously, the values for each virtual resources of VMs as well as its utilization can vary from its specified values (see Table \ref{table_inputs}).

Horizontal elasticity is considered in Table \ref{table_workload} in order to efficiently attend the increasing demand of resources scaling up the number of VMs of $S_1$ from 2 (at $t=0$) to 3 (at $t=1$) and from 3 (at $t=1$) to 4 (at $t=2$). The number of VMs scales down from 4 (at $t=3$) to 2 (at $t=4$), assuming a decreasing demand of resources. Additionally, vertical elasticity for processing and memory resources could be observed in the workload trace of Table \ref{table_workload} from $t=3$ to $t=4$, where $Vcpu_{111}''$ and $Vcpu_{121}''$ decrease from 8 [ECU] to 4 [ECU] and $Vram_{111}''$ and $Vram_{121}''$ decrease from 16 [GB] to 8 [GB], assuming a decreasing demand of resources. Vertical elasticity could also be applied to other resources as described in Section \ref{environments}.

Finally, server and network resources utilization change dynamically in VMs from $t=0$ to $t=4$, representing important data for CSPs in order to apply a safe overbooking of both server and network resources (see Table \ref{table_workload}). At $t=0$, it can be seen a high utilization of both processing and memory resource, representing a possible alarm for scaling up the number of VMs (horizontal elasticity) as can be observed in $t=1$. Low utilization of resources can be seen at $t=3$, representing an alarm for scaling down both the number of VMs (horizontal elasticity) as well as hardware configuration of each VM (vertical elasticity) as can be seen at $t=4$. 

\section{Conclusions and Future Work}
\label{conclusions}
Based on an universe of 64 studied publications carefully chosen as explained in Section II, this work extended the taxonomy presented in \cite{lopez2015} focusing on dynamic (online) formulations of the VMP problem from the providers' perspective complementing a previous work of the authors \cite{lopez2015} and proposed a novel taxonomy in order to understand possible challenges for Cloud Service Providers (CSPs) in dynamic environments to efficiently attend customers' request for virtual resources, based on the most relevant dynamic parameters studied so far in the VMP literature.

This work identified that resource capacities of VMs (associated to vertical elasticity), number of VMs of a cloud service (associated to horizontal elasticity) and utilization of resources of VMs (related to overbooking) are the most relevant dynamic parameters in literature. Consequently, dynamic environments for online formulations of the provider-oriented VMP problem were classified by one of the following classification criteria: (1) elasticity and (2) overbooking. First, dynamic environments could be formulated considering one of the following elasticity values: no elasticity, horizontal elasticity, vertical elasticity or both horizontal and vertical elasticity. Additionally, identified dynamic environments may also
consider one of the following overbooking values: no overbooking, server resources overbooking, network resources overbooking or both server and network overbooking.

Based on the combinations of the possible values of the classification criteria (elasticity and overbooking), the proposed taxonomy identified 16 different possible environments (see Figure \ref{table_taxonomy}), characterizing each environment with particular mathematical notation of time variables (see Table \ref{table_dynamic_notation}).

The proposed taxonomy showed that research of online formulations of the provider-oriented VMP problem has been mainly studied in Environment (0,0) and (0,1) with 39\% and 50\% of the studied articles respectively. Other briefly studied environments are Environment (0,2), (2,0), (2,1) and (3,0). Several research opportunities for unexplored environments were identified (see Figure \ref{table_taxonomy}). For example, no paper was found studying horizontal elasticity alone, even more, joint network and server overbooking is still a field with no published paper. Considering both types of elasticity and both types of overbooking represent the most advanced environment identified in this work: Environment (3,3). CSPs efficiently solving formulations of the VMP problem in this complex (3,3) dynamic environment will represent a considerable advance on this research area and its cloud datacenters will be able to scale according to trending types of requirements with sufficient flexibility. A recommended path for future work is exploring and addressing challenges of particular environments identified as research opportunities before considering this advanced and complete (3,3) dynamic environment for solving the provider-oriented VMP problem.

At the time of this writing, the authors are already working on extending identified environments to consider dynamic level of SLAs and dynamic revenue for executing VMs, just to cite a few characteristics to be included. Additionally, other environments could be studied considering dynamic electricity costs or pricing schemes in federated clouds, among others.

}

\bibliographystyle{IEEEtranS}
\bibliography{Taxonomy}

\begin{thebibliography}{10}
\providecommand{\url}[1]{#1}
\csname url@samestyle\endcsname
\providecommand{\newblock}{\relax}
\providecommand{\bibinfo}[2]{#2}
\providecommand{\BIBentrySTDinterwordspacing}{\spaceskip=0pt\relax}
\providecommand{\BIBentryALTinterwordstretchfactor}{4}
\providecommand{\BIBentryALTinterwordspacing}{\spaceskip=\fontdimen2\font plus
\BIBentryALTinterwordstretchfactor\fontdimen3\font minus
  \fontdimen4\font\relax}
\providecommand{\BIBforeignlanguage}[2]{{%
\expandafter\ifx\csname l@#1\endcsname\relax
\typeout{** WARNING: IEEEtranS.bst: No hyphenation pattern has been}%
\typeout{** loaded for the language `#1'. Using the pattern for}%
\typeout{** the default language instead.}%
\else
\language=\csname l@#1\endcsname
\fi
#2}}
\providecommand{\BIBdecl}{\relax}
\BIBdecl

\bibitem{alicherry2013optimizing}
M.~Alicherry and T.~Lakshman, ``Optimizing data access latencies in cloud
  systems by intelligent virtual machine placement,'' in \emph{INFOCOM, 2013
  Proceedings IEEE}.\hskip 1em plus 0.5em minus 0.4em\relax IEEE, 2013, pp.
  647--655.

\bibitem{anand2013virtual}
A.~Anand, J.~Lakshmi, and S.~Nandy, ``Virtual machine placement optimization
  supporting performance slas,'' in \emph{Cloud Computing Technology and
  Science (CloudCom), 2013 IEEE 5th International Conference on}, vol.~1.\hskip
  1em plus 0.5em minus 0.4em\relax IEEE, 2013, pp. 298--305.

\bibitem{armbrust2009m}
M.~Armbrust, O.~Fox, R.~Griffith, A.~D. Joseph, Y.~Katz, A.~Konwinski, G.~Lee,
  D.~Patterson, A.~Rabkin, I.~Stoica \emph{et~al.}, ``Above the clouds: A
  berkeley view of cloud computing,'' \emph{University of California, Berkeley,
  Tech. Rep}, 2009.

\bibitem{beloglazov2012energy}
A.~Beloglazov, J.~Abawajy, and R.~Buyya, ``Energy-aware resource allocation
  heuristics for efficient management of data centers for cloud computing,''
  \emph{Future Generation Computer Systems}, vol.~28, no.~5, pp. 755--768,
  2012.

\bibitem{Biran2012}
O.~Biran, A.~Corradi, M.~Fanelli, L.~Foschini, A.~Nus, D.~Raz, and E.~Silvera,
  ``A stable network-aware vm placement for cloud systems,'' in
  \emph{Proceedings of the 2012 12th IEEE/ACM International Symposium on
  Cluster, Cloud and Grid Computing (ccgrid 2012)}.\hskip 1em plus 0.5em minus
  0.4em\relax IEEE Computer Society, 2012, pp. 498--506.

\bibitem{buyya2008market}
R.~Buyya, C.~S. Yeo, and S.~Venugopal, ``Market-oriented cloud computing:
  Vision, hype, and reality for delivering it services as computing
  utilities,'' in \emph{High Performance Computing and Communications, 2008.
  HPCC'08. 10th IEEE International Conference on}, 2008.

\bibitem{Calcavecchia2012}
N.~M. Calcavecchia, O.~Biran, E.~Hadad, and Y.~Moatti, ``Vm placement
  strategies for cloud scenarios,'' in \emph{Cloud Computing (CLOUD), 2012 IEEE
  5th International Conference on}.\hskip 1em plus 0.5em minus 0.4em\relax
  IEEE, 2012, pp. 852--859.

\bibitem{cao2014energy}
Z.~Cao and S.~Dong, ``An energy-aware heuristic framework for virtual machine
  consolidation in cloud computing,'' \emph{The Journal of Supercomputing}, pp.
  1--23, 2014.

\bibitem{Chang2013}
D.~Chang, G.~Xu, L.~Hu, and K.~Yang, ``A network-aware virtual machine
  placement algorithm in mobile cloud computing environment,'' in
  \emph{Wireless Communications and Networking Conference Workshops (WCNCW),
  2013 IEEE}.\hskip 1em plus 0.5em minus 0.4em\relax IEEE, 2013, pp. 117--122.

\bibitem{chen2013intelligent}
K.-y. Chen, Y.~Xu, K.~Xi, and H.~J. Chao, ``Intelligent virtual machine
  placement for cost efficiency in geo-distributed cloud systems,'' in
  \emph{Communications (ICC), 2013 IEEE International Conference on}.\hskip 1em
  plus 0.5em minus 0.4em\relax IEEE, 2013, pp. 3498--3503.

\bibitem{dalvandi2013time}
A.~Dalvandi, M.~Gurusamy, and K.~C. Chua, ``Time-aware vm-placement and routing
  with bandwidth guarantees in green cloud data centers,'' in \emph{Cloud
  Computing Technology and Science (CloudCom), 2013 IEEE 5th International
  Conference on}, vol.~1.\hskip 1em plus 0.5em minus 0.4em\relax IEEE, 2013,
  pp. 212--217.

\bibitem{dang2013higher}
H.~T. Dang and F.~Hermenier, ``Higher sla satisfaction in datacenters with
  continuous vm placement constraints,'' in \emph{Proceedings of the 9th
  Workshop on Hot Topics in Dependable Systems}.\hskip 1em plus 0.5em minus
  0.4em\relax ACM, 2013, p.~1.

\bibitem{Dias2012}
D.~S. Dias and L.~H.~M. Costa, ``Online traffic-aware virtual machine placement
  in data center networks,'' in \emph{Global Information Infrastructure and
  Networking Symposium (GIIS), 2012}.\hskip 1em plus 0.5em minus 0.4em\relax
  IEEE, 2012, pp. 1--8.

\bibitem{Do2011}
A.~V. Do, J.~Chen, C.~Wang, Y.~C. Lee, A.~Y. Zomaya, and B.~B. Zhou,
  ``Profiling applications for virtual machine placement in clouds,'' in
  \emph{Cloud Computing (CLOUD), 2011 IEEE International Conference on}.\hskip
  1em plus 0.5em minus 0.4em\relax IEEE, 2011, pp. 660--667.

\bibitem{Dong2013}
D.~Dong and J.~Herbert, ``Energy efficient vm placement supported by data
  analytic service,'' in \emph{Cluster, Cloud and Grid Computing (CCGrid), 2013
  13th IEEE/ACM International Symposium on}.\hskip 1em plus 0.5em minus
  0.4em\relax IEEE, 2013, pp. 648--655.

\bibitem{dong2013virtual}
J.~Dong, H.~Wang, X.~Jin, Y.~Li, P.~Zhang, and S.~Cheng, ``Virtual machine
  placement for improving energy efficiency and network performance in iaas
  cloud,'' in \emph{Distributed Computing Systems Workshops (ICDCSW), 2013 IEEE
  33rd International Conference on}.\hskip 1em plus 0.5em minus 0.4em\relax
  IEEE, 2013, pp. 238--243.

\bibitem{Dupont2012}
C.~Dupont, G.~Giuliani, F.~Hermenier, T.~Schulze, and A.~Somov, ``An energy
  aware framework for virtual machine placement in cloud federated data
  centres,'' in \emph{Future Energy Systems: Where Energy, Computing and
  Communication Meet (e-Energy), 2012 Third International Conference on}.\hskip
  1em plus 0.5em minus 0.4em\relax IEEE, 2012, pp. 1--10.

\bibitem{fang2013power}
S.~Fang, R.~Kanagavelu, B.-S. Lee, C.~H. Foh, and K.~M.~M. Aung,
  ``Power-efficient virtual machine placement and migration in data centers,''
  in \emph{Green Computing and Communications (GreenCom), 2013 IEEE and
  Internet of Things (iThings/CPSCom), IEEE International Conference on and
  IEEE Cyber, Physical and Social Computing}.\hskip 1em plus 0.5em minus
  0.4em\relax IEEE, 2013, pp. 1408--1413.

\bibitem{fang2013vmplanner}
W.~Fang, X.~Liang, S.~Li, L.~Chiaraviglio, and N.~Xiong, ``Vmplanner:
  Optimizing virtual machine placement and traffic flow routing to reduce
  network power costs in cloud data centers,'' \emph{Computer Networks},
  vol.~57, no.~1, pp. 179--196, 2013.

\bibitem{Ferreto2011}
T.~Ferreto, C.~A. De~Rose, and H.-U. Heiss, ``Maximum migration time guarantees
  in dynamic server consolidation for virtualized data centers,'' in
  \emph{Euro-Par 2011 Parallel Processing}.\hskip 1em plus 0.5em minus
  0.4em\relax Springer, 2011, pp. 443--454.

\bibitem{Ferreto2011a}
T.~C. Ferreto, M.~A. Netto, R.~N. Calheiros, and C.~A. De~Rose, ``Server
  consolidation with migration control for virtualized data centers,''
  \emph{Future Generation Computer Systems}, vol.~27, no.~8, pp. 1027--1034,
  2011.

\bibitem{gahlawat2014survey}
M.~Gahlawat and P.~Sharma, ``Survey of virtual machine placement in federated
  clouds,'' in \emph{Advance Computing Conference (IACC), 2014 IEEE
  International}.\hskip 1em plus 0.5em minus 0.4em\relax IEEE, 2014, pp.
  735--738.

\bibitem{georgiou2013exploiting}
S.~Georgiou, K.~Tsakalozos, and A.~Delis, ``Exploiting network-topology
  awareness for vm placement in iaas clouds,'' in \emph{Cloud and Green
  Computing (CGC), 2013 Third International Conference on}.\hskip 1em plus
  0.5em minus 0.4em\relax IEEE, 2013, pp. 151--158.

\bibitem{Goudarzi2012}
H.~Goudarzi and M.~Pedram, ``Energy-efficient virtual machine replication and
  placement in a cloud computing system,'' in \emph{Cloud Computing (CLOUD),
  2012 IEEE 5th International Conference on}.\hskip 1em plus 0.5em minus
  0.4em\relax IEEE, 2012, pp. 750--757.

\bibitem{guo2013shadow}
Y.~Guo, A.~L. Stolyar, and A.~Walid, ``Shadow-routing based dynamic algorithms
  for virtual machine placement in a network cloud,'' in \emph{INFOCOM, 2013
  Proceedings IEEE}.\hskip 1em plus 0.5em minus 0.4em\relax IEEE, 2013, pp.
  620--628.

\bibitem{gupta2013hpc}
A.~Gupta, L.~V. Kal{\'e}, D.~Milojicic, P.~Faraboschi, and S.~M. Balle,
  ``Hpc-aware vm placement in infrastructure clouds,'' in \emph{Cloud
  Engineering (IC2E), 2013 IEEE International Conference on}.\hskip 1em plus
  0.5em minus 0.4em\relax IEEE, 2013, pp. 11--20.

\bibitem{Gupta2012}
A.~Gupta, D.~Milojicic, and L.~V. Kal{\'e}, ``Optimizing vm placement for hpc
  in the cloud,'' in \emph{Proceedings of the 2012 workshop on Cloud services,
  federation, and the 8th open cirrus summit}.\hskip 1em plus 0.5em minus
  0.4em\relax ACM, 2012, pp. 1--6.

\bibitem{Ho2011}
Y.~Ho, P.~Liu, and J.-J. Wu, ``Server consolidation algorithms with bounded
  migration cost and performance guarantees in cloud computing,'' in
  \emph{Utility and Cloud Computing (UCC), 2011 Fourth IEEE International
  Conference on}.\hskip 1em plus 0.5em minus 0.4em\relax IEEE, 2011, pp.
  154--161.

\bibitem{hoeflin2012quantifying}
D.~Hoeflin and P.~Reeser, ``Quantifying the performance impact of overbooking
  virtualized resources,'' in \emph{Communications (ICC), 2012 IEEE
  International Conference on}.\hskip 1em plus 0.5em minus 0.4em\relax IEEE,
  2012, pp. 5523--5527.

\bibitem{hong2013qoe}
H.-J. Hong, D.-Y. Chen, C.-Y. Huang, K.-T. Chen, and C.-H. Hsu, ``Qoe-aware
  virtual machine placement for cloud games,'' in \emph{Network and Systems
  Support for Games (NetGames), 2013 12th Annual Workshop on}.\hskip 1em plus
  0.5em minus 0.4em\relax IEEE, 2013, pp. 1--2.

\bibitem{Huang2013}
W.~Huang, X.~Li, and Z.~Qian, ``An energy efficient virtual machine placement
  algorithm with balanced resource utilization,'' in \emph{Innovative Mobile
  and Internet Services in Ubiquitous Computing (IMIS), 2013 Seventh
  International Conference on}.\hskip 1em plus 0.5em minus 0.4em\relax IEEE,
  2013, pp. 313--319.

\bibitem{Huang2012b}
Z.~Huang and D.~H. Tsang, ``Sla guaranteed virtual machine consolidation for
  computing clouds,'' in \emph{Communications (ICC), 2012 IEEE International
  Conference on}.\hskip 1em plus 0.5em minus 0.4em\relax IEEE, 2012, pp.
  1314--1319.

\bibitem{Huang2012}
Z.~Huang, D.~H. Tsang, and J.~She, ``A virtual machine consolidation framework
  for mapreduce enabled computing clouds,'' in \emph{Proceedings of the 24th
  International Teletraffic Congress}.\hskip 1em plus 0.5em minus 0.4em\relax
  International Teletraffic Congress, 2012, p.~26.

\bibitem{hwang2013hierarchical}
I.~Hwang and M.~Pedram, ``Hierarchical virtual machine consolidation in a cloud
  computing system,'' in \emph{Cloud Computing (CLOUD), 2013 IEEE Sixth
  International Conference on}.\hskip 1em plus 0.5em minus 0.4em\relax IEEE,
  2013, pp. 196--203.

\bibitem{Jiang2012a}
J.~W. Jiang, T.~Lan, S.~Ha, M.~Chen, and M.~Chiang, ``Joint vm placement and
  routing for data center traffic engineering,'' in \emph{INFOCOM, 2012
  Proceedings IEEE}.\hskip 1em plus 0.5em minus 0.4em\relax IEEE, 2012, pp.
  2876--2880.

\bibitem{Jin2013}
H.~Jin, H.~Qin, S.~Wu, and X.~Guo, ``Ccap: A cache contention-aware virtual
  machine placement approach for hpc cloud,'' \emph{International Journal of
  Parallel Programming}, pp. 1--18, 2013.

\bibitem{Jin2012}
H.~Jin, D.~Pan, J.~Xu, and N.~Pissinou, ``Efficient vm placement with multiple
  deterministic and stochastic resources in data centers,'' in \emph{Global
  Communications Conference (GLOBECOM), 2012 IEEE}.\hskip 1em plus 0.5em minus
  0.4em\relax IEEE, 2012, pp. 2505--2510.

\bibitem{kakadia2013network}
D.~Kakadia, N.~Kopri, and V.~Varma, ``Network-aware virtual machine
  consolidation for large data centers,'' in \emph{Proceedings of the Third
  International Workshop on Network-Aware Data Management}.\hskip 1em plus
  0.5em minus 0.4em\relax ACM, 2013, p.~6.

\bibitem{Kantarci2012}
B.~Kantarci, L.~Foschini, A.~Corradi, and H.~T. Mouftah, ``Inter-and-intra data
  center vm-placement for energy-efficient large-scale cloud systems,'' in
  \emph{Globecom Workshops (GC Wkshps), 2012 IEEE}.\hskip 1em plus 0.5em minus
  0.4em\relax IEEE, 2012, pp. 708--713.

\bibitem{Kord2013}
N.~Kord and H.~Haghighi, ``An energy-efficient approach for virtual machine
  placement in cloud based data centers,'' in \emph{Information and Knowledge
  Technology (IKT), 2013 5th Conference on}.\hskip 1em plus 0.5em minus
  0.4em\relax IEEE, 2013, pp. 44--49.

\bibitem{Le2011}
K.~Le, R.~Bianchini, J.~Zhang, Y.~Jaluria, J.~Meng, and T.~D. Nguyen,
  ``Reducing electricity cost through virtual machine placement in high
  performance computing clouds,'' in \emph{Proceedings of 2011 International
  Conference for High Performance Computing, Networking, Storage and
  Analysis}.\hskip 1em plus 0.5em minus 0.4em\relax ACM, 2011, p.~22.

\bibitem{Li2013a}
K.~Li, J.~Wu, and A.~Blaisse, ``Elasticity-aware virtual machine placement for
  cloud datacenters,'' in \emph{Cloud Networking (CloudNet), 2013 IEEE 2nd
  International Conference on}.\hskip 1em plus 0.5em minus 0.4em\relax IEEE,
  2013, pp. 99--107.

\bibitem{li2013migration}
K.~Li, H.~Zheng, and J.~Wu, ``Migration-based virtual machine placement in
  cloud systems,'' in \emph{Cloud Networking (CloudNet), 2013 IEEE 2nd
  International Conference on}.\hskip 1em plus 0.5em minus 0.4em\relax IEEE,
  2013, pp. 83--90.

\bibitem{Li2012a}
W.~Li, J.~Tordsson, and E.~Elmroth, ``Virtual machine placement for predictable
  and time-constrained peak loads,'' in \emph{Economics of Grids, Clouds,
  Systems, and Services}.\hskip 1em plus 0.5em minus 0.4em\relax Springer,
  2012, pp. 120--134.

\bibitem{li2013energy}
X.~Li, Z.~Qian, S.~Lu, and J.~Wu, ``Energy efficient virtual machine placement
  algorithm with balanced and improved resource utilization in a data center,''
  \emph{Mathematical and Computer Modelling}, vol.~58, no.~5, pp. 1222--1235,
  2013.

\bibitem{Lin2012}
J.-W. Lin and C.-H. Chen, ``Interference-aware virtual machine placement in
  cloud computing systems,'' in \emph{Computer \& Information Science (ICCIS),
  2012 International Conference on}, vol.~2.\hskip 1em plus 0.5em minus
  0.4em\relax IEEE, 2012, pp. 598--603.

\bibitem{lopez2015b}
F.~L{\'o}pez-Pires and B.~Bar{\'a}n, ``A many-objective optimization framework
  for virtualized datacenters,'' in \emph{Proceedings of the 2015 5th
  International Conference on Cloud Computing and Service Science}, 2015, pp.
  439--450.

\bibitem{lopez2014survey}
\BIBentryALTinterwordspacing
F.~L\'opez-Pires and B.~Bar\'an, ``Virtual machine placement literature
  review,'' Polytechnic School, National University of Asunci\'on, Tech. Rep.,
  2015. [Online]. Available: \url{http://arxiv.org/abs/1506.01509}
\BIBentrySTDinterwordspacing

\bibitem{lopez2015}
F.~L{\'o}pez-Pires and B.~Bar{\'a}n, ``A virtual machine placement taxonomy,''
  in \emph{Proceedings of the 2015 IEEE/ACM 15th International Symposium on
  Cluster, Cloud and Grid Computing}.\hskip 1em plus 0.5em minus 0.4em\relax
  IEEE Computer Society, 2015.

\bibitem{lopez2013virtual}
F.~L{\'o}pez-Pires, E.~Melgarejo, and B.~Bar{\'a}n, ``Virtual machine
  placement. a multi-objective approach,'' in \emph{Computing Conference
  (CLEI), 2013 XXXIX Latin American}.\hskip 1em plus 0.5em minus 0.4em\relax
  IEEE, 2013, pp. 1--8.

\bibitem{lu2013qos}
K.~Lu, R.~Yahyapour, P.~Wieder, C.~Kotsokalis, E.~Yaqub, and A.~I. Jehangiri,
  ``Qos-aware vm placement in multi-domain service level agreements
  scenarios,'' in \emph{Cloud Computing (CLOUD), 2013 IEEE Sixth International
  Conference on}.\hskip 1em plus 0.5em minus 0.4em\relax IEEE, 2013, pp.
  661--668.

\bibitem{mills2011comparing}
K.~Mills, J.~Filliben, and C.~Dabrowski, ``Comparing vm-placement algorithms
  for on-demand clouds,'' in \emph{Cloud Computing Technology and Science
  (CloudCom), 2011 IEEE Third International Conference on}.\hskip 1em plus
  0.5em minus 0.4em\relax IEEE, 2011, pp. 91--98.

\bibitem{Mishra2011}
M.~Mishra and A.~Sahoo, ``On theory of vm placement: Anomalies in existing
  methodologies and their mitigation using a novel vector based approach,'' in
  \emph{Cloud Computing (CLOUD), 2011 IEEE International Conference on}.\hskip
  1em plus 0.5em minus 0.4em\relax IEEE, 2011, pp. 275--282.

\bibitem{moreno2013improved}
I.~S. Moreno, R.~Yang, J.~Xu, and T.~Wo, ``Improved energy-efficiency in cloud
  datacenters with interference-aware virtual machine placement,'' in
  \emph{Autonomous Decentralized Systems (ISADS), 2013 IEEE Eleventh
  International Symposium on}.\hskip 1em plus 0.5em minus 0.4em\relax IEEE,
  2013, pp. 1--8.

\bibitem{ortigoza2015}
\BIBentryALTinterwordspacing
J.~Ortigoza, F.~L\'opez-Pires, and B.~Bar\'an, ``Workload trace generation for
  dynamic environments in cloud computing,'' Polytechnic School, National
  University of Asunci\'on, Tech. Rep., 2015. [Online]. Available:
  \url{http://arxiv.org/abs/1507.00090}
\BIBentrySTDinterwordspacing

\bibitem{Piao2010}
J.~T. Piao and J.~Yan, ``A network-aware virtual machine placement and
  migration approach in cloud computing,'' in \emph{Grid and Cooperative
  Computing (GCC), 2010 9th International Conference on}.\hskip 1em plus 0.5em
  minus 0.4em\relax IEEE, 2010, pp. 87--92.

\bibitem{prevost2013optimal}
J.~J. Prevost, K.~Nagothu, B.~Kelley, and M.~Jamshidi, ``Optimal update
  frequency model for physical machine state change and virtual machine
  placement in the cloud,'' in \emph{System of Systems Engineering (SoSE), 2013
  8th International Conference on}.\hskip 1em plus 0.5em minus 0.4em\relax
  IEEE, 2013, pp. 159--164.

\bibitem{ribas2013pbfvmc}
B.~C. Ribas, R.~M. Suguimoto, R.~A. Montano, F.~Silva, and M.~Castilho,
  ``Pbfvmc: A new pseudo-boolean formulation to virtual-machine
  consolidation,'' in \emph{Intelligent Systems (BRACIS), 2013 Brazilian
  Conference on}.\hskip 1em plus 0.5em minus 0.4em\relax IEEE, 2013, pp.
  201--206.

\bibitem{Ribas2012}
B.~C. Ribas, R.~M. Suguimoto, R.~A. Montano, F.~Silva, L.~de~Bona, and M.~A.
  Castilho, ``On modelling virtual machine consolidation to pseudo-boolean
  constraints,'' in \emph{Advances in Artificial Intelligence--IBERAMIA
  2012}.\hskip 1em plus 0.5em minus 0.4em\relax Springer, 2012, pp. 361--370.

\bibitem{salimian2013survey}
L.~Salimian and F.~Safi, ``Survey of energy efficient data centers in cloud
  computing,'' in \emph{Proceedings of the 2013 IEEE/ACM 6th International
  Conference on Utility and Cloud Computing}.\hskip 1em plus 0.5em minus
  0.4em\relax IEEE Computer Society, 2013, pp. 369--374.

\bibitem{Sato2013}
K.~Sato, M.~Samejima, and N.~Komoda, ``Dynamic optimization of virtual machine
  placement by resource usage prediction,'' in \emph{Industrial Informatics
  (INDIN), 2013 11th IEEE International Conference on}.\hskip 1em plus 0.5em
  minus 0.4em\relax IEEE, 2013, pp. 86--91.

\bibitem{shi2013provisioning}
L.~Shi, B.~Butler, D.~Botvich, and B.~Jennings, ``Provisioning of requests for
  virtual machine sets with placement constraints in iaas clouds,'' in
  \emph{Integrated Network Management (IM 2013), 2013 IFIP/IEEE International
  Symposium on}.\hskip 1em plus 0.5em minus 0.4em\relax IEEE, 2013, pp.
  499--505.

\bibitem{Shi2011}
W.~Shi and B.~Hong, ``Towards profitable virtual machine placement in the data
  center,'' in \emph{Utility and Cloud Computing (UCC), 2011 Fourth IEEE
  International Conference on}.\hskip 1em plus 0.5em minus 0.4em\relax IEEE,
  2011, pp. 138--145.

\bibitem{Shigeta2013}
S.~Shigeta, H.~Yamashima, T.~Doi, T.~Kawai, and K.~Fukui, ``Design and
  implementation of a multi-objective optimization mechanism for virtual
  machine placement in cloud computing data center,'' in \emph{Cloud
  Computing}.\hskip 1em plus 0.5em minus 0.4em\relax Springer, 2013, pp.
  21--31.

\bibitem{singh2013reduce}
N.~A. Singh and M.~Hemalatha, ``Reduce energy consumption through virtual
  machine placement in cloud data centre,'' in \emph{Mining Intelligence and
  Knowledge Exploration}.\hskip 1em plus 0.5em minus 0.4em\relax Springer,
  2013, pp. 466--474.

\bibitem{song2014optimization}
F.~Song, D.~Huang, H.~Zhou, H.~Zhang, and I.~You, ``An optimization-based
  scheme for efficient virtual machine placement,'' \emph{International Journal
  of Parallel Programming}, vol.~42, no.~5, pp. 853--872, 2014.

\bibitem{soundararajan2010challenges}
V.~Soundararajan and K.~Govil, ``Challenges in building scalable virtualized
  datacenter management,'' \emph{ACM SIGOPS Operating Systems Review}, vol.~44,
  no.~4, pp. 95--102, 2010.

\bibitem{Speitkamp2010}
B.~Speitkamp and M.~Bichler, ``A mathematical programming approach for server
  consolidation problems in virtualized data centers,'' \emph{Services
  Computing, IEEE Transactions on}, vol.~3, no.~4, pp. 266--278, 2010.

\bibitem{tsai2013prevent}
M.-H. Tsai, J.~Chou, and J.~Chen, ``Prevent vm migration in virtualized
  clusters via deadline driven placement policy,'' in \emph{Cloud Computing
  Technology and Science (CloudCom), 2013 IEEE 5th International Conference
  on}, vol.~1.\hskip 1em plus 0.5em minus 0.4em\relax IEEE, 2013, pp. 599--606.

\bibitem{Tsakalozos2011}
K.~Tsakalozos, M.~Roussopoulos, and A.~Delis, ``Vm placement in non-homogeneous
  iaas-clouds,'' in \emph{Service-Oriented Computing}.\hskip 1em plus 0.5em
  minus 0.4em\relax Springer, 2011, pp. 172--187.

\bibitem{wang2013particle}
S.~Wang, Z.~Liu, Z.~Zheng, Q.~Sun, and F.~Yang, ``Particle swarm optimization
  for energy-aware virtual machine placement optimization in virtualized data
  centers,'' in \emph{Parallel and Distributed Systems (ICPADS), 2013
  International Conference on}.\hskip 1em plus 0.5em minus 0.4em\relax IEEE,
  2013, pp. 102--109.

\bibitem{wang2014eqvmp}
S.-H. Wang, P.~P.-W. Huang, C.~H.-P. Wen, and L.-C. Wang, ``Eqvmp:
  Energy-efficient and qos-aware virtual machine placement for software defined
  datacenter networks,'' in \emph{Information Networking (ICOIN), 2014
  International Conference on}.\hskip 1em plus 0.5em minus 0.4em\relax IEEE,
  2014, pp. 220--225.

\bibitem{Wang2012}
W.~Wang, H.~Chen, and X.~Chen, ``An availability-aware virtual machine
  placement approach for dynamic scaling of cloud applications,'' in
  \emph{Ubiquitous Intelligence \& Computing and 9th International Conference
  on Autonomic \& Trusted Computing (UIC/ATC), 2012 9th International
  Conference on}.\hskip 1em plus 0.5em minus 0.4em\relax IEEE, 2012, pp.
  509--516.

\bibitem{Wu2012b}
J.-J. Wu, P.~Liu, and J.-S. Yang, ``Workload characteristics-aware virtual
  machine consolidation algorithms,'' in \emph{Proceedings of the 2012 IEEE 4th
  International Conference on Cloud Computing Technology and Science
  (CloudCom)}.\hskip 1em plus 0.5em minus 0.4em\relax IEEE Computer Society,
  2012, pp. 42--49.

\bibitem{Zamanifar2012}
K.~Zamanifar, N.~Nasri, and M.~Nadimi-Shahraki, ``Data-aware virtual machine
  placement and rate allocation in cloud environment,'' in \emph{Advanced
  Computing \& Communication Technologies (ACCT), 2012 Second International
  Conference on}.\hskip 1em plus 0.5em minus 0.4em\relax IEEE, 2012, pp.
  357--360.

\bibitem{Zhang2013}
X.~Zhang, Y.~Zhang, X.~Chen, K.~Liu, G.~Huang, and J.~Zhan, ``A
  relationship-based vm placement framework of cloud environment,'' in
  \emph{Proceedings of the 2013 IEEE 37th Annual Computer Software and
  Applications Conference}.\hskip 1em plus 0.5em minus 0.4em\relax IEEE
  Computer Society, 2013, pp. 124--133.

\bibitem{zhang2014dynamic}
X.~Zhang, Q.~Yue, and Z.~He, ``Dynamic energy-efficient virtual machine
  placement optimization for virtualized clouds,'' in \emph{Proceedings of the
  2013 International Conference on Electrical and Information Technologies for
  Rail Transportation (EITRT2013)-Volume II}.\hskip 1em plus 0.5em minus
  0.4em\relax Springer, 2014, pp. 439--448.

\end{thebibliography}

\end{document}